\documentclass[12pt,nofootinbib]{revtex4-2}
\usepackage{amsmath,amssymb}
\newcommand{\ket}[1]{|#1 \rangle}
\newcommand{\CO}{{\cal O}}
\newcommand{\vev}[1]{\langle #1 \rangle}
\newcommand{\BR}{{\mathbb R}}

\begin{document}
\title{Operator Product Expansions and recoil}

\author{David Berenstein, Ruwanmali Bernadette de Zoysa}
\address{Department of Physics, University of California, Santa Barbara, CA 93106}
\begin{abstract}
Some issues of recoil effects in AdS/CFT are studied from the point of view of OPE expansions for generalized free fields. We show that the conformal group structure encodes
the relative energies and momenta at a collision center. This is done by being careful with the analysis of Clebsch-Gordan coefficients for an $SL(2)$ subalgebra of the
 conformal group. The collision fraction of kinetic energy carried by the particles is derived  from a  probability distribution that arises from these coefficients. We specifically identify a precise statement of when recoil of a heavy particle in AdS can be ignored:  the maximum probability is for the heavy particle to be in its ground state.
 We also argue how a notion of reduced mass appears in these collisions, in the limit where the particles are moving slowly with respect to each other. This controls the
 notion of the impact parameter of the collision. 
 \end{abstract}

\maketitle

\section{ Two intuitions collide}

In the AdS/CFT correspondence \cite{Maldacena:1997re, Witten:1998qj,Gubser:1998bc}, the gravity intuition is supposed to be equivalent to conformal field theory intuition, once we have  a dictionary that properly identifies how these two are supposed to match. 

One intuition, from certain bound state  problems, is that if we have a light particle bound to a heavy particle, we can assume that the center of mass is located at the position of the heavy particle. Another, one could say {\em fancier}, way to restate this intuition is that we can think of a heavy particle as a background field that has spontaneously broken translation invariance. 
This is how we would deal with solitons in field theory for example.  
The light field is then expanded relative to the heavy particle background. This is also how we deal with the hydrogen atom on a first approximation: we neglect the motion of the proton.

If we have spontaneous symmetry breaking,  there is a Goldstone boson associated to moving the heavy particle around. The Goldstone boson mechanics is used to restore translation invariance of the whole system.
The Goldstone mode can in principle receive small corrections from the light particle, but they are suppressed. As such, one does not make a big error in neglecting them on a first pass and we basically say that this approximation neglects recoil effects (this is one form of back-reaction). 

We can apply this same intuition to a black hole in AdS, or some other heavy object in the center of global AdS, like a D-brane for example. In this intuition, the symmetry group that is spontaneously broken is the conformal group. A Goldstone boson should appear that restores the symmetry. This does indeed happen, but because the commutation relations of the spontaneously broken generators with the Hamiltonian don't vanish  \footnote{This Hamitonian is the generator of global time translations, which is identified with the radial time quantization in CFT.}, the modes are massive (this is important in the discussion of the difference between mass and dimension for particles in $AdS$ \cite{Berenstein:2002ke}).

These modes generate the descendants of the representation of the conformal group associated to the heavy object. With a light field excitation added to the problem, the Goldstone mode should still be attached to the heavy object and may receive small corrections due to the light excitation.

Now, let us start with a different intuition from quantum field theory, associated to the use of the symbol $\partial_\mu^{\leftrightarrow}$. If we have two scalar fields $\phi_1, \phi_2$, the symbol is used as the following definition 
\begin{equation}
\phi_1 \partial_\mu^{\leftrightarrow} \phi_2\equiv \phi_1 \partial_{\mu} \phi_2-\phi_2\partial_\mu \phi_1
\end{equation}
and the definition can be iterated. For example, in a Feynman rule of a heavy particle $T$ of spin $s$ decaying into $\phi_1, \phi_2$, we would write a term in the effective Lagrangian of the form
\begin{equation}
\lambda T \cdot \phi_2 (\partial^{\leftrightarrow})^s \phi_1
\end{equation}
where all the tensor index labels are suppressed. If we are on-shell in the center of mass of particle $T$, $T$ has a transverse polarization. The components of momentum that appear in the symbol are the spacelike difference of momenta $\vec p_1-\vec p_2$ to the n-th power. In the center of mass $\vec p_1= -\vec p_2$, so the symbol can be ambiguous. 
What is important is that the combination is defined so that $\phi_2 (\partial^{\leftrightarrow})^s \phi_1$ is {\em not a total derivative}. If it were, we could integrate by parts and use transversality of the polarization of $T$ to argue that the Feynman rule for decay vanishes.
In this intuition, $\phi_1, \phi_2$ enter symmetrically even if they have a different mass.  Also, implicitly, the operator made of $\phi_1, \phi_2$ has spin $s$ so the indices 
in the derivatives are not contracted. 

Now, let us address the use of the symbol in conformal field theory. If we have a free field theory, it appears the same way as above when we have two free scalar operators.
For the symbol to make some sense more generally, we need to have a way to relate composite fields made of two operators to their na\"\i ve product and derivatives. A simple example where this occurs is in the presence of generalized free fields. These are ubiquitous in the AdS/CFT correspondence. The basic idea is that when we quantize the various modes around the vacuum in the large $N$ limit, interactions between particles are suppressed by $1/N$. This is also known as large $N$ factorization. 
We can assume that the spectrum is a  free Fock space of states for small energies above the vacuum. This is usually the set of fluctuations around the vacuum geometry, which can be a Fock space of string states and not just supergravity fluctuations.

Assume therefore that we have two different  primary operators $\CO_1$, $\CO_2$ (of dimension $\Delta_1, \Delta_2$ respectively) that act as generalized free fields. In that case, the correlation functions involving $\CO_1,\CO_2$ factorize. This is actually the definition of what generalized free fields do, so that a four point function like the one below is written as a product of known two point functions
\begin{equation}
\langle \CO_1(z_1)\CO_2(z_2)\CO_1(z_3)\CO_1(z_4)\rangle= \langle \CO_1(z_1)\CO_1(z_3)\rangle  \langle \CO_2(z_2)\CO_2(z_4)\rangle \label{eq:fact}
\end{equation}
What is important is that not only is the product $\CO_1(x)\CO_2(x)$ non singular (we just take limits), but that we can use Taylor series in $z_2-z_1$ to extract  composites with derivatives. The products are automatically normal ordered. 

The operator product expansion of $\CO_1$ and $\CO_2$ is given schematically by
\begin{equation}
\CO_1(z_1)\CO_2(z_2) \sim \sum f_{123}\frac 1{|z_1-z_2|^{\Delta_1+\Delta_2-\Delta_3}} \CO_3(z_1)\label{eq:OPE}
\end{equation}
where the sum is over all possible primaries $\CO_3$ and the $f_{123}$ are OPE coefficients.
Because the correlation function is known from \eqref{eq:fact}, we can actually write the list of possible $\CO_3$ representations of the conformal group that appear in the right hand side: they are the list of operators described above with derivatives acting on $\CO_1,\CO_2$.  
 The objects $\CO_3(z_1) $ are primaries and are  sometimes written also as $\CO_1 (\partial^\leftrightarrow)^n \CO_2$ (see for example \cite{Heemskerk:2009pn} , which also includes a longer list if we allow contractions between the indices). Because they are primaries, 
they are {\em not descendants}. Descendants are total derivatives after all. The use of the symbol is a shorthand to say that the primaries appearing in the OPE are primaries made of derivatives of $\CO_1$ and $\CO_2$, whose dimension is $\Delta_3= \Delta_1+\Delta_2+n$. 

The puzzle arises in how do we reconcile the use of the symbol  $(\partial^\leftrightarrow)^n$ in this example with the heavy versus light intuition above. Namely, the na\"\i ve use of the symbol is symmetric in the entries, whereas the  heavy versus light intuition treats the two entries in a distinct manner.
In this setup, this occurs when $\Delta_1 \gg \Delta_2$. The point is that only when the fields have the same dimension is the answer symmetric in the two operators. The na\"\i ve symbol of iterated derivatives does not describe the correct primary field.

Beyond generalized free fields, some theorems in the asymptotic behavior of the OPEs \cite{Fitzpatrick:2014vua,Fitzpatrick:2012yx,Komargodski:2012ek}  argue that 
operators that appear in the right hand side of the OPE start behaving as products of generalized free fields at large dimension  and fixed twist. More precisely,
the dimensions of operators for large spin $J$ on the right hand side of \eqref{eq:OPE} asymptotically have the dimension of the operators in the generalized free field setup, namely $\Delta_3\sim \Delta_1+\Delta_2+n+O(1/J)$ and $n=J+f$ where $f$ is finite. The twist is $\tau_3=\Delta_3-J\sim \Delta_1+\Delta_2+f+O(1/J)$, which is finite.
 One uses the same   symbol $\CO_3\sim\CO_1 (\partial^\leftrightarrow)^n \CO_2$ to indicate that for very large $n$ (the dimension of the primaries)  the right hand side will generically be very similar to the result of generalized free fields between the two operators. Hence the symbol $\CO_1 (\partial^\leftrightarrow)^n \CO_2$ is allowed to be used in the right hand side with small corrections on the dimension of $\CO_3$ as a short hand for this intuition. One could also expect that the OPE coefficient is similar to the implicit OPE coefficient in \eqref{eq:OPE}. This is also a way to start addressing the problem of scattering of particles in AdS at large $N$, in a way that should be complementary to the Mellin amplitude program initiated in  \cite{Penedones:2010ue,Fitzpatrick:2011ia}. Such a calculation of binding energies would be analogous to 
problems at finite volume in QFT that are used to describe scattering phases  \cite{Luscher:1990ux}.

Our goal is to explain when we need to take into account recoil and when it can be neglected. Either when we use the intuition of the generalized free field OPE or the  gravity intuition (kinematics of collisions)  with a massive object in the center.
 For example,  there is a distinction between an electron and a proton moving very slowly with respect to each other in a collision and a very relativistic  collision between them. In the first case, the proton is carrying most of the energy in the center of mass and the center of mass frame is very close to the frame where the proton is at rest. In this same case, the electron carried most of the kinetic energy.
  In the second case we can assume both are effectively massless and each carries half the energy of the collision in the center of mass thereby  appearing on the same footing. There is a transition between the two regimes and our goal is to understand how this transition works from kinematics of the OPE expansion of generalized free fields. Overall, we are understanding the technical details of kinematics of the conformal group in the generalized free field setup and situations which can be well approximated by such a setup.

 The paper is organized as follows. In section \ref{sec:CG} we explain how the problem we need to study is related to tensor products of representations for $SL(2,\BR)$ (basically, addition of angular momentum for $SL(2,\BR)$). 
 In section\ref{sec:ref} we compute carefully how the probabilistic averages of fraction of energy carried by each constituent can be calculated in various limits and how these coincide with the notions of momentum kinematics of the center of mass frame for a collision
 in AdS. In section \ref{sec:ang} we define the notion of impact parameter in AdS (point of closest approach in the center of mass). We show that this leads to the notion of reduced mass for non-relativistic collisions. in the presence of recoil, we also show the asymptotics for
 maximal angular momentum configurations and how the distance is effective larger with recoil included than without, we also explain some details of how back-reaction geometrically determine some effects that have been discovered in the bootstrap program. We then conclude.

\section{$SL(2)$ Clebsch-Gordan coefficients}\label{sec:CG}

Before we compute our main result, which is dependent on the computation of $SL(2,{\mathbb R})$ Clebsch-Gordan coefficients, we need some preliminary notation to normalize conventions with respect to 
conformal field theory.
The $SL(2)$ Lie algebra has three generators, $L_+, L-, L_0$. The commutation relations are 
\begin{eqnarray}
\ [L_0,L_+] &=& L_+\\
\ [L_0, L_-]&=& - L_-\\
\ [L_-,L_+]&=& 2 L_0
\end{eqnarray}
For unitary representations, by convention $L_0$ is hermitian, and $L_-^\dagger = L_+$.
A lowest weight state or primary, is a state $\ket{\Delta}$ such that $L_- \ket \Delta =0$.
It is said to have weight $\Delta$, where $L_0\ket \Delta = \Delta \ket \Delta$ is the eigenvalue of the state with respect to $L_0$. 
Sometimes $L_0$ is also called the energy operator.  The lowest state minimizes the energy in a given representation and completely characterizes it.
The representations we need have $L_0$ bounded from below, but not from above.

In conformal field theory, we have usually a longer list of generators $\Delta, K_\mu, P_\mu, M_{\mu\nu} $. The operator $\Delta$ is the generator of dilations, $P_\mu$ are translations and $M_{\mu\nu}$ are
rotations. The operators $K_\mu$ generate special conformal transformations. 

The operators $P_\mu$ act on all operators as derivatives
\begin{equation}
[P_\mu, \CO(x)]= i \partial_\mu \CO(x)
\end{equation}
and if $\CO$ is a primary field, we have that 
\begin{equation}
[K_\mu, \CO(0)]=0
\end{equation}
The descendants are generated by $P_\mu$ acting multiple times on the primary operators.
Via the operator state correspondence, we send primary (and descendant) fields to states for the conformal field theory on the cylinder
\begin{eqnarray}
\CO(0)&\to& \ket \CO\\
\ [P_\mu, \CO] &\to& P_\mu\ket \CO
\end{eqnarray}
and the unitarity condition is $K_\mu= P_\mu^\dagger$ on the cylinder. 

If we choose a single variable $x$ in the Euclidean plane, we call $P_x\sim \partial_x\equiv \partial$ and we identify it with $L_+$. The adjoint of the $P_x$ operator on the cylinder (this generator is  $K_x$) is matched with $L_-$. The object that plays the role of $2 L_0$ is the commutator $[L_-, L_+]$ which is given in this case by $2\Delta$. Sometimes it is convenient to use a holomorphic $z$ rather than real $x$, and the the role of $L_0$ becomes a combination of dilatation and angular momentum in the $z$ plane. This is not important for us right now, but later it will become important in the paper. The point is that we still have and $SL(2,{\mathbb R})$ algebra. The representation theory that we need does not change
on these modifications.

What we have been able to do is find a subalgebra of the conformal group in higher dimensions that coincides with $SL(2,{\mathbb R})$, which we will simply call $SL(2)$ in the rest of the paper.

In the problem of generalized free fields, a primary field $\CO_1$ of dimension $\Delta_1$ will have energy $\Delta_1$ in the cylinder. The descendants under repeated action of $L_{+}$,  are given by 
$\partial^n\CO$ and will have dimension $\Delta_1+n$. The inner product determining the norm is the Zamolodchikov metric. It coincides with a natural notion of inner product of the Hilbert space in the cylinder.
For product composite operators in generalized free fields $\CO_1\CO_2$, we have that the energies add $\Delta_{tot}= \Delta_1+\Delta_2$. Adding derivatives  is done by the Leibnitz rule. A quick exercise shows that in this case the lists of operators appearing in the list of possible operators can have derivatives distributed in arbitrary amounts between 
the two operators. This is exactly the list of elements of the product of representations of both conformal fields in the cylinder Hilbert space.
The decomposition of the resulting objects is the same as the decomposition into irreducible representations of the product representation in $SL(2)$.

This is very similar to the theory of addition of angular momentum and this is the point. Our computations are about recoil effects between fields that don't interact, so the notion of conserved charges adds between  different pieces: there is no binding energy between the parts.
 To determine the primary states at each dimension $\Delta+n$ we need to solve the problem of finding the primaries (lowest weight states) in the product of representations of $SL(2)$. This is a special case of Clebsch-Gordan coefficients for products of $SL(2)$ representations. 

A primary state $\ket \Delta$ is  chosen to be normalized.  Unitarity (positivity of the inner product)  requires that $\Delta > 0$.
The representation with $\Delta=0$ is trivial.
Descendants are obtained by acting repeatedly with $L_+$ on $\ket \Delta$.
We will label them as $\ket{\Delta, n}$ where $n$ is the number of $L_+$ acting on $\ket \Delta$. These states have energy $\Delta +n$.
If we choose these to be normalized with unit norm, then one can easily show that $L_+$ acts as follows
\begin{equation}
L_+ \ket{\Delta; n}= \sqrt{ (n+1)(n+2 \Delta) }\ket{\Delta; n+1} \label{eq:sllp}
\end{equation}
This is similar to the raising and lowering operators of angular momentum, where
\begin{equation}
J^{SU(2)}_+ \ket{j; m}= \sqrt{j(j+1)-m(m+1)}\ket{j; m+1}\label{eq:angmlp}
\end{equation}
To get from \eqref{eq:angmlp} to \eqref{eq:sllp}, we can analytically continue by taking $j=  -\Delta, m= (\Delta+n)$ and changing the total sign inside the square root \footnote{The usual theory of angular momentum starts with a highest weight state. For $SL(2)$ we start with a lowest weight state,. In $SU(2)$ such lowest weight state would have angular momentum $-j$ rather than $j$.}.
By unitarity, we get that 
\begin{equation}
L_- \ket{\Delta; n}= \sqrt{ (n)(n-1+2 \Delta) }\ket{\Delta; n-1} \label{eq:sllp2}
\end{equation}

Our goal now is to understand the most general state in the product representation wit fixed dimension $\Delta_1+\Delta_2+n$. These are given by
\begin{equation}
\hbox{Span}( \ket{\Delta_1, n_1}\ket{\Delta_2, n-n_1})
\end{equation}
Notice that these states are orthonormalized. If we do a calculation with $\partial^{k} \CO$ instead, these do not have unit norm. To compensate for this choice of non normalized basis the coefficients appearing in \eqref{eq:sllp},\eqref{eq:sllp2} would be different. This modifies how one would compute in intermediate steps. 
We now need to solve the equation that determines primary states
\begin{eqnarray}
L_-\left(\sum a_{n_1}  \ket{\Delta_1, n_1}\ket{\Delta_2, n-n_1} \right)&=& 0\nonumber\\
 \sum a_{n_1} (L_- \ket{\Delta_1, n_1})\ket{\Delta_2, n-n_1} + a_{n_1}  \ket{\Delta_1, n_1}L_-\ket{\Delta_2, n-n_1}&=& 0\label{eq:lws}
\end{eqnarray}
This gives a simple recursion
\begin{eqnarray}
a_{n_1+1}\sqrt{(n_1+1)(n_1+2\Delta_1)}    + a_{n_1} \sqrt{(n-n_1)(n-n_1-1+2 \Delta_2)}=0\label{eq:recrel}
\end{eqnarray}
which is the coefficient of $\ket{\Delta_1, n_1}\ket{\Delta_2, n-n_1-1}$ in the expression \eqref{eq:lws}.
We can solve the recursion in a straightforward way
\begin{equation}
a_{\ell} =  (-1)^{\ell}\left(\prod_{s=0}^{\ell-1} \frac{(n-s)(n-s-1+2 \Delta_2)}{(s+1)(s+2 \Delta_1)}\right)^{1/2}
 a_0\end{equation}
 or equivalently
 \begin{equation}
 a_{\ell}= (-1)^\ell \left(\frac{(n)!}{\ell! (n-\ell)!} \frac{\Gamma(n+2 \Delta_2)\Gamma(2 \Delta_1)}{\Gamma(n+2 \Delta_2-\ell)\Gamma(\ell+2 \Delta_1)}\right)^{1/2} a_0
 \end{equation}
As a special case
\begin{equation}
a_n= (-1)^n  \left( \frac{\Gamma(n+2 \Delta_2)\Gamma(2 \Delta_1)}{\Gamma(2 \Delta_2)\Gamma(n+2 \Delta_1)}\right)^{1/2} a_0\label{eq:recsol}
\end{equation}
and this can be used to show that there is symmetry under $a_\ell\rightarrow a_{n-\ell}$, $\Delta_1\leftrightarrow \Delta_2$.
To complete the calculation of the Clebsch-Gordan coefficients we would also need to orthonormalize the answer.
This requires that 
\begin{equation}
\sum_{\ell=0}^n |a_\ell|^2=1 
\end{equation}
but that detail is not needed in what follows.

\section{Recoil effects}\label{sec:ref}

Let us start our discussion of recoil with two particles colliding in flat space, of masses $m_1> m_2$. The center of mass frame is the frame where the kinetic energy takes its minimum value. The net spatial momentum is zero in the center of mass frame, so that $\vec p= \vec p_2=-\vec p_1$. The total energy is 
\begin{equation}
E= \sqrt{\vec p^2 + m_2^2}+\sqrt{\vec p^2 + m_1^2}\label{eq:kinetic}
\end{equation}
and the kinetic energy is
\begin{equation}
K=\sqrt{|\vec p|^2 + m_2^2}-m_2+\sqrt{|\vec p|^2 + m_1^2}-m_1\to \frac{|\vec p|^2}{2 m_2}+\frac{|\vec p|^2}{2 m_1} \label{eq:nonrel}
\end{equation}
where the rightmost equation is the limit of $|\vec p|$ small. In this limit, the fraction of the kinetic energy carried by the heavy particle versus the light one is fixed and related to the ratio of the masses. The kinetic energy is also suppressed relative to the rest energy. 

Most of the kinetic energy is in the light particle, but the ratio is related to the rest masses.
Consider now the large $|p|\gg m_1, m_2$ regime. In that case, the kinetic energy is the larger fraction of the total energy and both contributions are comparable to each other.
If the masses are very different, there is an intermediate regime where the momentum is much larger than the smaller mass, but much smaller than the larger one. In that regime,
which we dub the semi-relativistic limit, 
the light particle can be taken as essentially massless, so the kinetic energy is roughly linear in $|\vec p|$ and the kinetic energies  start becoming comparable to each other when $|p|\simeq m_2$, when the heavy particle is starting to become relativistic. 
In this case we have that 
\begin{equation}
K\sim |\vec p| + \frac{|\vec p|^2}{2 m_1}\label{eq:semir}
\end{equation}
We want to understand a similar notion that is encoded  in \eqref{eq:recsol}. Unlike flat space, where the energy is a continuous variable, in $AdS$ the energy is quantized.
We need to understand the physics with this in mind.

First, we will take the heavy particle to be associated to $\Delta_1$ and the lighter one to $\Delta_2$. The total energy will be $\Delta_1+\Delta_2+n$. The notion of kinetic energy 
should be equal to $n$: we subtract the  energy of the system with the two particles at rest (each in their ground state). 
We want to know which of the two particles is mostly responsible for carrying the kinetic energy of the configuration. 

The main idea is to understand that in the computation in the previous section, the amplitude squared $|a_s|^2$ is (proportional to) the probability that the heavy particle carries kinetic energy $s$. We will say recoil can be neglected if $|a_1|^2 <|a_0|^2$, that is, if the probability of being in the ground state for the heavy particle is the largest one of them all.

This condition is that 
\begin{equation}
r=\frac{a_1^2}{a_0^2} =  \frac{n(n-1+2 \Delta_2)}{2\Delta_1}<1\label{eq:nonrelcond}
\end{equation}
It is easy to show that if this is true, then $|a_{s+2}/{a_{s+1}}|^2<|a_{s+1}/{a_s}|^2$. This follows straightforwardly from the recursion relation \eqref{eq:recrel}. We find that
\begin{equation}
\left(\frac{a_{s+1}^2}{a_s^2}\right)= \frac{(n-s)(n-s-1+2 \Delta_2)}{(s+1)(s+2\Delta_1)}
\end{equation}
so that if the equation \eqref{eq:nonrelcond} above is satisfied, the numerator decreases and the denominator increases: the fraction stays smaller than the one before. 

The simplest limit we can take is where we first take $\Delta_2\gg n $ so that the transition value can be estimated  $n\simeq \Delta_1/\Delta_2$. Recoil becomes important relatively quickly if $\Delta_1, \Delta_2$ are quantities of the same order of magnitude. On the other hand, the value of  $n$ at which the transition occurs can be quite large if there is a large difference  in the dimensions $\Delta_1, \Delta_2$.
More generally, if the value of $n$ at which we transition is much larger than $2 \Delta_2$, we find that $n\sim \sqrt{2 \Delta_1}$.

In these cases the fraction of the kinetic energy carried  by the heavy particle is not zero. Let us now look at the simplest case $n\ll \Delta_2 
< \Delta_1 $, but where we might allow ourselves to take $r>1$. 
In this limit the ratios of the Gamma functions and factorials simplify,  so  we have that 
\begin{equation}
\frac{a_\ell^2}{a^2_0}\sim \frac{n^\ell}{\ell!} \frac{(2 \Delta_2)^\ell}{(2 \Delta_1)^\ell} = \frac{r^{\ell}}{\ell!}
\end{equation}
We found that the energy carried by the heavy particle is approximately a Poisson distribution with  mean given by
\begin{equation}
\vev{\ell} = r \sim n \frac{\Delta_2}{\Delta_1}
\end{equation}
the fraction of the kinetic energy carried by the heavy particle is 
\begin{equation}
f= \frac{\vev{\ell} }{n}= \frac{\Delta_2}{\Delta_1}
\end{equation}
very similar to the non-relativistic fraction in \eqref{eq:nonrel} if we identify the mass $m_1, m_2\sim\Delta_1,  \Delta_2$ respectively.  The energy carried in recoil by the particle is fractional and not an integer. This is because we are averaging over states with the total energy fixed.

 This is just like we would expect in the AdS/CFT dictionary between masses of particles in $AdS$ and the dimension of operators in the dual conformal field theory when these are large \cite{Witten:1998qj} (the shift of $d/2$ in this limit can be attributed to zero point energy of the particle at the bottom of AdS, where it behaves like a harmonic oscillator \cite{Berenstein:2002ke} ). We clearly see that  if $r<1$ we can really neglect recoil. The heavy particle is essentially at rest, with a fraction of one quantum of excitation allowed. 

For the next limit, we take $\Delta_2 \ll n \ll \Delta_1$. A similar simplification occurs in the gamma functions, where now we have that $n$ dominates over $\Delta_1$. In this case we find that
\begin{equation}
\frac{a^2_\ell}{a^2_0}\sim \frac{n^\ell}{n!} \frac{n^\ell}{(2\Delta_1)^\ell}
\end{equation}
and again we find an approximate  Poisson distribution with
\begin{equation}
\vev{\ell}=r=\frac{a_1^2}{a_0^2} \sim \frac{n^2}{2 \Delta_1}\ll n 
\end{equation}
this is again very similar to the behavior in the semi-relativistic regime \eqref{eq:semir}, if we realize that  we should identify $|\vec p|\sim n$. The same intuition says that recoil is important only if $r>1$, so that $n> \sqrt{2 \Delta_1}$.

Finally, we take the limit where $n$ is larger than everything else. This limit is handled by expanding the Gamma functions in $|a_\ell|^2$ in the Stirling approximation.
There we find that 
\begin{eqnarray}
\log(a_\ell^2) &\sim& C - (\ell \log(\ell)-\ell)-((n-\ell) \log(n-\ell) -(n-\ell))\\
&&
-\left[( \ell+2\Delta_1 -1)\log( \ell+2\Delta_1 -1)-(\ell+2\Delta_1 -1)) \right]
\\
&&-\left[  ( n-\ell+2\Delta_2 -1)\log( ( n-\ell+2\Delta_2 -1)- ( n-\ell+2\Delta_2 -1))\   \right]
\end{eqnarray}
where $C$ is a constant. 

Maximizing the amplitude with respect to $\ell$ we find that the most likely value of $\ell$ satisfies
\begin{equation}
\log(\ell) -\log(n-\ell) +\log( 2 \Delta_1+\ell) -\log(2 \Delta_2+n-\ell) =0
\end{equation}
where we can ignore the shifts by $1$ in the logs of the Gamma functions. 
The distribution of $\ell$ is effectively Gaussian near this maximum.
Equivalently, after some manipulation, 
\begin{equation}
\vev\ell \sim \frac 12 n \left( 1-\frac{\Delta_1-\Delta_2} {\Delta_1+\Delta_2+n}  \right)\label{eq:vevl}
\end{equation}

We need to compare this with the relativistic limit of equation \eqref{eq:kinetic}. In that limit we have that
\begin{equation}
K \sim 2 |p|
-m_1-m_2 
\end{equation}
With our identifications $m_1, m_2\sim \Delta_1,\Delta_2$, $n\sim K$, we would get that 
\begin{equation}
2 |p| = n+ \Delta_1 +\Delta_2
\end{equation}
and the kinetic energy of the first particle is
\begin{equation}
K_1= |p|-m_1 \sim \frac{n}{2} -\frac{\Delta_1-\Delta_2}{2}
\end{equation}
which is the correct limit in \eqref{eq:vevl} when $n\gg \Delta_1,\Delta_2$. Indeed, \eqref{eq:vevl} also covers the non-relatvistic and  semi-relativistic limit we derived before  so long as  $\ell\gg1$ so that Stirlings approximation is valid for moderate values of $\ell$ and $\Delta_2\ll \Delta_1$. 
When the situation is more symmetric, and $\Delta_1$ and $\Delta_2$  are of the same order of magnitude,and when $n\ll \Delta_2$  we find 
\begin{equation}
\ell \sim  n\frac{ \Delta_2}{\Delta_2+\Delta_1}\label{eq:non-relfrac}
\end{equation}
so that the ratio of kinetic energies is exactly the ratio of the masses, as expected in non-relativistic collisions.

Basically, in all cases we get a match with the kinetic theory of head-on collisions in flat space and the structure of the primaries ${\cal O}_1(\partial^\leftrightarrow)^n {\cal O}_2$.
This kinetic theory of the conformal group for generalized free fields does not depend on the AdS/CFT correspondence. 

Finally, let us state differently what the no recoil statement implies. It states that to a very good approximation, the objects
\begin{equation}
{\cal O}_1 \partial^n {\cal O}_2\label{eq:allin2}
\end{equation}
are  all primary operators. To a very good approximation, the $r-th$ descendants are of the form
\begin{equation}
(\partial^r {\cal  O}_1)  \partial^n {\cal O}_2
\end{equation}
This is exactly the intuition  of the heavy object spontaneously breaking the conformal group, and that the group is restored by a Goldstone mode that is attached to 
the heavy object. The $\partial^r$ is acting exactly the same as multiple excitations (exactly $r$ of them) of said Goldstone mode. 

At very high energies the two objects appear more symmetrically.

\section{Angular momentum and AdS geometry}\label{sec:ang}

So far we have discussed how the conformal group tensor product of representations  encodes the (momentum) kinematics of collisions at the origin of $AdS$.
To discuss the notion of impact parameter, we have to go beyond momentum and back to the notion of  position. This requires the AdS geometry, so that we can talk about position and not just momentum.
It is not obvious that we should be able to do that in general beyond holography. On the other hand, since $AdS$ is a homogeneous space for the conformal group, it is the natural arena to discuss 
the position kinematics anyhow. This would be similar to how one thinks of flat space as a group quotient of the Poincar\'e group by the Lorentz group (as is familiar in superspace discussions, see for example \cite{Wess:1992cp} ). We will take this philosophy in mind when we discuss distances between objects. Essentially, the presentations of the conformal group are a quantization of geometric motion in the AdS geometry for objects of different masses.

Let us say we have a primary scaler operator ${\cal O}$ of dimension $\Delta\gg 1$, and the descendant $\partial^n {\cal O}$. If the primary is at the center of  AdS,
 depending on the spin of the descendant, we will occupy different semiclassical trajectories in the bulk of AdS. We thus need to keep track of spin more carefully than we have been doing so far. We need to reintroduce spacetime indices in the derivatives.
 \begin{equation}
 \ket{\Delta;n, s}\sim \left(\partial_{\mu_1}\dots \partial_{\mu_s} (\partial_\mu\partial^\mu)^{(n-s)/2)} -\hbox{traces}\right){\cal O}
 \end{equation}
where the traces guarantee that the $\mu$ indices appear in a symmetric traceless combination (see for example \cite{Heemskerk:2009pn}). This is simplest to accomplish if all the $\mu_{1, \dots s}$ are the same and correspond to derivatives with respect to a complex coordinate. Thus, we look at the simplified states
 \begin{equation}
 \ket{\Delta;n, s}\sim \left(\partial_z^s (\partial_\mu\partial^\mu)^{(n-s)/2)} \right){\cal O}
 \end{equation}
To understand the geometry, we need to understand the classical geodesics of global AdS. We do this following the presentation in \cite{Berenstein:2019tcs}.
For that , we  need the metric of $AdS$ in global coordinates
\begin{equation}
ds^2 = -\cosh(\rho)^2 dt^2 + d\rho^2 + \sinh(\rho)^2 d\Omega^2
\end{equation}
and $\rho$ is the radial direction. Measuring distances to the origin (at constant time) results in a distance equal to $\rho$. That is why this systems of coordinates is convenient for us.

For the case $s=0$, which is the most similar case  to what we computed in the section \ref{sec:CG}, a particle in the $s$ wave follows a radial infalling trajectory.
The total energy of the particle corresponds to a radial turning point $\rho^*$, determined by  $E= m\cosh(\rho^*)$. When we have a system of two particles in their center of mass, and with relative angular momentum equal to zero, they are at the same point $\rho=0$ simultaneously.  Technically, because  we have declared ourselves to be in the lowest weight state of a representation, we can assume that the geometric notion of center of mass is that of a point particle at the origin of $AdS$. The center of mass is not wobbling.

That also means that the two particles each reach their turning point simultaneously, and the maximum distance between them is
\begin{equation}
d_{\max} = \rho_1^*+\rho_2^*
\end{equation} 
In the absence of back-reaction we take $\rho_1^*\sim 0$, so the maximum distance is 
\begin{equation}
d_{\max}\sim \cosh^{-1}\left(\frac{E_2}{m_2} \right)= \cosh^{-1}\left(1+\frac{n}{\Delta_2}\right)
\end{equation}
where we have used that the energy of the light particle is $E_2= \Delta_2+n$ and $m\sim \Delta_2$. Basically, this is how we translate to gravity that the primary operator in question is $\CO_1\partial^n \CO_2$ as described in equation \eqref{eq:allin2}.
In the small oscillation regime (non-relativistic limit for $\Delta_2$), where $d_{max}$ is small, we get  that 
\begin{equation}
\frac {d_{max}^2}{2} = \frac{n}{\Delta_2}
\end{equation}
This is exactly the image of the bottom of AdS being like a harmonic oscillator potential.
When we have recoil, from  \eqref{eq:non-relfrac} we would get that
\begin{equation}
d^2_{max,2}= 2 \frac{\Delta_{1}}{\Delta_1+\Delta_2}  \frac{n}{\Delta_2}
\end{equation}
and similar for $d_1$. The total 
\begin{equation}
d_{tot,max}= d_{max,2}+d_{max,1} = \sqrt{2 \frac {n}{\Delta_1+\Delta_2} }\left[\sqrt{\frac{\Delta_2}{\Delta_1}}+\sqrt{\frac{\Delta_1}{\Delta_2}}\right]
\end{equation}
with a bit of manipulation, we find that this is equal to 
\begin{equation}
d_{tot,max}= \sqrt{2 \frac {n}{\Delta_1+\Delta_2} } \left[  \frac{\Delta_2+\Delta_1}{\sqrt{\Delta_1\Delta_2}}\right]=\sqrt{\frac{2n}{\Delta_{red}}}
\end{equation}
where the quantity $\Delta_{red}$ is given by
\begin{equation}
\Delta_{red}= \frac{\Delta_1\Delta_2}{\Delta_1+\Delta_2}=\frac{ \Delta_2}{1+\frac{\Delta_2}{\Delta_1}}
\end{equation}
is a notion of reduced mass (reduced dimension in this case). We then write  
\begin{equation}
\frac{d_{tot,max}^2}{2}= \frac{n}{\Delta_{red}} \label{eq:dmax}
\end{equation}
We still have a harmonic oscillator behavior, with the same frequency as the one of $AdS$. Clearly, since $\Delta_{red}<\Delta_2$, the effect of recoil is to have the particles farther away that we would otherwise naively believe.
Alternatively, it is as if we had a single mass $\Delta_{red}$ oscillating at the bottom of AdS. Just like $d_{tot, max}$ is larger, so is the relative velocity at the point of collision.

The same can be done with the impact parameter in this non-relativistic regime, where we have some non-trivial angular momentum. In that case there is both an outer edge of the orbit and an inner edge.
What we see is that the position kinematics of the origin of AdS gives similar notions for non-relativistic collisions in flat space, where the physics is dominated by forces and an angular repulsion controlled by the reduced mass of a system. The details don't  change enough to require a different derivation.
The only extra observation we require is that there is a one parameter family of embeddings in the conformal group, given by
\begin{equation}
L^+ \propto p_x+ i \beta p_y
\end{equation}
In this embedding $L_0 \sim (1+\beta^2) \Delta+ 2 \beta M_{xy}$. Basically, the algebra we used doesn't change, but the identification of $L^{\pm}$ changes. 
There is similarly an orthogonal component to $L_0$ that annihilates $L^+$. This one is used to show that the fraction of spin carried by the heavy object is identical to the fraction of energy carried by the object.

Incidentally, these are the types of primaries needed to compute anomalous dimensions in quantum field theory, where both objects have the dimension of free fields. For more information see the review \cite{Beisert:2010jr} and references therein.

From a direct analysis, the notion of impact parameter would be given, in the non-relativistic regime, by an effective potential where we have no radial motion 
\begin{equation}n=
V_{eff}=\frac{\Delta_{red}r^2}2+\frac{s^2}{2\Delta_{red} r^2}
\end{equation}
The perihelion (in the non-relativistic limit) is the smallest root of the effective potential. This root is given by
\begin{equation}
\Delta_{red} r_{min}^2 = n-\sqrt{n^2-s^2}
\end{equation}
whereas the aphelion is at
\begin{equation}
\Delta_{red} r_{max}^2 = n+\sqrt{n^2-s^2}
\end{equation}
When $s\to0$, the smallest root goes to zero and the large root goes to the expression \eqref{eq:dmax}. 
We also see that a real solution requires $n\geq |s|$, which  can be interpreted as a unitarity bound (in the quantum system the spin is less that $n$ by algebraic construction). Violating it would require us to have states below the minimal energy of each representation. These would violate unitarity.

For more general situations, we would turn to equation (60) in \cite{Berenstein:2019tcs}, where the maximum and minimum radii can be determined for orbits in global $AdS$. The results are not enlightening. What does proceed, however, is that the fraction of angular momentum and kinetic energy carried by each particle is the same.
 This is because we can consider a different 
embedding of $SL(2)$ into the conformal algebra. One can see that at maximal spin, the representation theory we studied does not change at all, but we have that the excess energy is the spin $n = s$ (this would be at fixed twist equal to $\Delta_1+\Delta_2$). Again, the effect of recoil is to have the particles further apart than without recoil effects taken into account. Both particles also reach their aphelion and perihelion in their  orbit at the same time $t$ in the coordinates we introduced, just like the non-relativistic limits. This is the 
notion of the center of mass not wobbling.

It is useful to study the simplest case where we have circular orbits. Then we have that
\begin{equation}
\cosh(\rho^*) ^2= \ell/\Delta+1
\end{equation} 
At very large angular momentum $\ell$ we get that
\begin{equation}
\rho^* \simeq \log(2\ell/\Delta)
\end{equation}
Without backreaction, we have that 
\begin{equation}
\rho^* \simeq \frac 12\log(n/\Delta_2)\sim \frac 12 \log(n)
\end{equation}
whereas with backreaction we get that
\begin{equation}
\rho_{tot}=\rho^*_1+\rho^*_2\sim \frac 12 \log(n/\Delta_1)+\frac 12 \log(n/\Delta_2)\sim \log(n)
\end{equation}
so that we are almost twice as far.

If we add interactions, that means that with recoil effects taken into account we expect slightly weaker interactions between the particles if they are dominated by the exchange of heavy particles (they are further apart after all). This is important to consider carefully when we are scattering from a  heavy target in holographic setups, like a black hole or a D-brane. The effects are exponential in the distance, times the mass of the exchange particle
\begin{eqnarray}
\Delta E \sim \exp(- m \rho_{tot})&\sim& \frac{1}{s^{m/2}} \hbox{ with no recoil}\\
&\sim&  \frac{1}{s^{m}} \hbox{ with recoil} 
\end{eqnarray}
where $s$ is the total maximal spin ($s=n$).

From the analysis, if D-branes have mass of order $N$ (as is typical in AdS), the recoil effects for a particle of mass of order one (strings) are of order $1/N$ (for example, the corrections in the reduced mass are of order  $\Delta_1/\Delta_2$ and these also control the notion of distance). This is just as expected: they are suppressed by the open string coupling, rather than the closed string coupling constant \footnote{ In principle one is able to compute them by virtual effects in flat space \cite{Berenstein:1996xk}.}. To understand recoil, one needs to introduce collective coordinates. This can be done with D-branes in some quantum field theory duals to AdS setups \cite{Berenstein:2013md}, although in that case the collective coordinate is internal to  additional coordinates of the manifold. The fact that we have a non-linear realization of the conformal group with Goldstone boson excitations would let us do the same, so long as we use coherent states in $SL(2)$. The recoil can be thought of as the virtual effects of the Goldstone boson dynamics (for more information see \cite{Gervais:1975yg}). 
These coherent states can be built as group rotations of ground states. This is a formal way of saying that $AdS$ is a group coset space of the conformal group. 
For example, one has 
\begin{equation}
AdS_{d+1} \equiv SO(d,2)/SO(d,1)
\end{equation}
This way of thinking has been used to write a superstring action in $AdS_5\times S^5$ \cite{Metsaev:1998it}. 

For black holes, with energy of order $N^2$, the recoil for light particles (strings) is of order $1/N^2$, the closed string coupling constant \cite{tHooft:1973alw}: we can consider them as gravitational back-reaction effects. Notice that in the asymptotic calculations in \cite{Fitzpatrick:2014vua} where the free field behavior start dominating, one has that the angular momentum $J$ is much larger than the dimensions of the  operators involved.
The corrections in dimension are of order $1/J^{\tau-1}$, where $\tau$ is the twist of the exchange operator.  For massive scalar field exchange this is roughly $\tau\sim m$. 
 By our results,  recoil needs to be taken into account. Without recoil effects, the results are slightly different (see for example \cite{Berenstein:2020vlp}), where the result is $1/J^{(\tau-1)/2}$. We see that apart\ from the offset, we get the exact scaling. 
  This can also be understood in the heavy/light bootstrap \cite{Li:2019zba}.
The difference can be completely ascribed to the fact that with recoil the  scattering object is farther from the black hole than we would otherwise believe.

\section{Conclusion}

In this paper we have studied the kinematics of tensor products of representations of the conformal group. This is a starting point for studying scattering in AdS in a way that is complementary to the Mellin space approach.

The basic idea is that there is a notion of center of mass frame in $AdS$. It is the frame where the energy is minimized within a given representation of the conformal group. This is equivalent to being in the lowest weight state (the primary state)
of a given conformal representation. When we apply this intuition, we find that in this center of mass frame we can recover the fractions of the energy carried by each object in a way that matches the kinematics of collisions in flat space.

We are also able to define an impact parameter in terms of the minimal distance between the two objects in the classical orbit that they perform. In the non-relativistic limit, we recover the notion of reduced mass (reduced dimension) and the impact parameter can be found easily from 
an effective harmonic oscillator potential with an angular momentum repulsion corresponding to the spin of the representation. 
It is always the case that recoil effects between a heavy and a light particle increases the effective distance between them. This explains a difference in scaling in the asymptotic limits of the bootstrap between the heavy-light approximation and the universal very high angular momentum approximation.
The effect is entirely kinematical.

In all of our calculations the local physics in AdS was used as a way to focus our intuition, but holography is not really required. The momentum space kinematic computations depend only on the Clebsch-Gordan coefficients of the conformal group tensor product of representations. 
These are independent of holography.
Similarly, for the notion of impact parameter, we can think of a background AdS space as a coset space of the conformal group: it can exist on its own without holography. If this is useful or not will depend on if local  physics on this geometry is a good approximation or not. 
\acknowledgements

D. B. would like to thank D.Grabovsky, Z. Li, J. Simon for discussions. The work of D.B. is supported in part by the Department of Energy under grant DE-SC 0011702.

\bibliography{refs}

\end{document}